\documentclass[journal]{IEEEtran}
\usepackage{amsmath,amssymb}
\usepackage{subfigure}
\usepackage{graphicx,graphics,color,psfrag}
\usepackage{cite,balance}
\usepackage{caption}
\allowdisplaybreaks
\usepackage{algorithm}
\usepackage{algorithmic}
\usepackage{accents}
\usepackage{amsthm}
\usepackage{bm}
\usepackage{url}
\usepackage[english]{babel}
\usepackage{multirow}
\usepackage{enumerate}
\usepackage{cases}
\usepackage{stfloats}
\usepackage{dsfont}
\usepackage{color,soul}
\usepackage{amsfonts}
\usepackage{cite,graphicx,amsmath,amssymb}
\usepackage{subfigure}
\usepackage{fancyhdr}
\usepackage{hhline}
\usepackage{graphicx,graphics}
\usepackage{array,color}
\usepackage{mathtools}
\usepackage{amsmath}

\newtheorem{lemma}{\emph{\underline{Lemma}}}

\newtheorem{proposition}{\emph{\underline{Proposition}}}

\newtheorem{remark}{\bf \emph{\underline{Remark}}}

\def\l{\left}
\def\r{\right}
\def\({\left(}
\def\){\right)}

\def\b0{{\mathbf{0}}}

\newcommand{\diag}{\mathrm{diag}}

\newcommand{\nn}{\nonumber}

\usepackage[top=0.65in, bottom=0.65in, left=0.66in, right=0.66in]{geometry}

\begin{document}
\captionsetup[figure]{name={Fig.}}

\title{\huge 
 Wireless Communication Aided by Intelligent\\ Reflecting Surface: Active or Passive?} 
\author{{Changsheng You,~\emph{Member,~IEEE}
	 and Rui Zhang,~\emph{Fellow,~IEEE}} 
	 \vspace{-18pt}
	   \thanks{\noindent This work is supported by MOE Singapore under Award T2EP50120-0024 and by ARTIC of National University of Singapore under Grant R-261-518-005-720.  (\emph{Corresponding author: Changsheng You}.)
	   
	   The authors are with the Department of Electrical and Computer Engineering, National University of Singapore, Singapore (Email: \{eleyouc, elezhang\}@nus.edu.sg). 
}}
\maketitle

\begin{abstract}
In this letter, we consider an intelligent reflecting surface (IRS)-aided wireless communication system, where an \emph{active} or \emph{passive} IRS is employed to assist the communication between an access point and a user.  First, we consider the downlink/uplink communication separately and optimize the IRS placement for rate maximization with an active or passive IRS. {\color{black}We show that given the same number of IRS reflecting elements,} the active IRS should be deployed closer to the receiver with the IRS's decreasing  amplification power; while in contrast, the passive IRS should be deployed near either the transmitter or receiver. Moreover, with optimized IRS placement, the passive IRS is shown to outperform its active counterpart when the number of reflecting elements is sufficiently large and/or the active-IRS amplification power is too small. Next, we optimize the IRS placement for both active and passive  IRSs to maximize the weighted sum-rate of uplink and downlink communications. We show that in this case, the passive IRS is more likely to achieve superior rate performance. This is because the optimal active-IRS placement needs to balance the rate performance in the uplink and downlink, while deploying the passive IRS near the transmitter or receiver is optimal regardless of the uplink or downlink.

\end{abstract}

\vspace{-5pt}
\section{Introduction}

Intelligent reflecting surface (IRS) has emerged as a promising technology to reconfigure the radio propagation environment in favor of wireless communications, by smartly tuning the reflection amplitude and/or phase-shift at each of its large number of  reflecting elements \cite{wu2021intelligent,qingqing2019towards,di2020smart}. Moreover, IRS is generally of light weight and profile, thus can be easily coated on environment objects to enhance the communication performance in different wireless systems (see, e.g., \cite{Wu2019TWC,you2019progressive,Pan2020Multicell,you2020deploy}).

Specifically, most of the existing works on IRS have considered the \emph{passive} IRS that reflects signals with passive loads (positive resistances) only. As such, it operates in full-duplex (FD) mode and is free of amplification/processing noise as well as self-interference, thus yielding higher spectral and energy efficiency as compared to the conventional active relay \cite{wu2021intelligent,di2020reconfigurable}. Nevertheless, the performance of passive-IRS aided system may be constrained by its high \emph{(product-distance)} path-loss \cite{wu2021intelligent}. This issue can be alleviated by installing a large number of passive reflecting elements to reap the square-order beamforming gain and/or deploying the passive IRS close to the transmitter and/or receiver to reduce the path-loss \cite{wu2021intelligent}.  In contrast, a new type of IRS, called \emph{active} IRS, has been recently proposed (see, e.g., \cite{long2021active,zhang2021active}) to overcome the practical  issue of passive IRS,  by amplifying the reflected signal with low-cost hardware. Generally speaking,  active IRS comprises a number of active reflecting elements, which are equipped with negative resistance components such as tunnel diode and negative impedance converter, thus enabled to reflect the incident signal with \emph{power amplification} \cite{lonvcar2020challenges,amato2018tunneling}. Different from the conventional amplify-and-forward (AF) relay that requires power-hungry radio frequency (RF) chains and orthogonal time/frequency resources to receive and transmit signals with power amplification, the active IRS directly reflects signals in an FD manner with low-power reflection-type amplifiers. Moreover, it was shown in \cite{long2021active,zhang2021active} that given the same IRS location, the active-IRS aided system tends to achieve a higher downlink rate than the  passive-IRS counterpart  due to the amplification gain, albeit at a modestly higher hardware and energy cost. However, the superiority of active IRS over passive IRS in terms of rate performance still remains unknown if their deployment is optimized, as well as when both the uplink and downlink communications are considered.

To address the above  questions, we consider in this letter an IRS-aided wireless communication system where a single-antenna access point (AP) communicates with a single-antenna user, aided by an active or passive IRS.  First, we consider the downlink/uplink  communication separately  and optimize the IRS placement for rate maximization with an active or passive IRS {\color{black}given the same budge on the number of IRS reflecting elements}. We show that the active IRS should be deployed closer to the receiver with the decreasing amplification power of the active IRS. This is in sharp contrast to the optimal placement of  the passive IRS, which is known to be near the transmitter or receiver to minimize the product-distance path-loss \cite{wu2021intelligent}. Moreover, with optimized IRS placement, the passive IRS is shown to even outperform its active counterpart when the number of reflecting elements is sufficiently large and/or the amplification power of the active IRS  is too small. Next, we optimize the IRS placement for both active and passive IRSs to maximize  the weighted sum-rate of uplink and downlink communications. We show that in this case, the passive IRS is more likely to achieve superior rate performance to the active IRS, because the optimal active-IRS placement needs to balance the rate performance in the uplink and downlink, while deploying the passive IRS near the transmitter or receiver is optimal for both the uplink and downlink.

\vspace{-5pt}
\section{System Model}\label{Sec:Model}

\begin{figure}[t]
\begin{center}
\includegraphics[width=8cm]{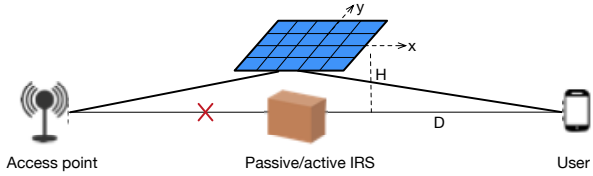}
\caption{IRS-aided wireless communication systems, where the IRS can be active or passive.}
\label{Fig:Syst}
\end{center}
\vspace{-5pt}
\end{figure}

Consider an IRS-aided wireless communication system shown in Fig.~\ref{Fig:Syst}, where
a single-antenna AP communicates with a single-antenna  user.\footnote{The obtained results can be extended to the case of multi-antenna AP by considering  the AP's transmit/receive beamforming.} We consider the challenging scenario where the AP-user direct link is severely blocked due to dense obstacles. Thus, an active or passive IRS with $N$ reflecting elements  is deployed  between them to assist their uplink and downlink communications. {\color{black}For simplicity, we assume that the IRS deployed at an altitude of $H$ with its half-space serving region towards the ground can establish line-of-sight (LoS) links with both the AP and user on the ground by proper deployment}.
 Let $D$, $x_{\rm AI}$, and $x_{\rm IU}$ denote  the horizontal distances between the AP and user, the AP and IRS, and the IRS and user, respectively, with $x_{\rm AI}+x_{\rm IU}\!=\!D$. Thus, the AP-IRS and IRS-user distances are given by $d_{\rm AI}\!=\!\sqrt{x_{\rm AI}^2+H^2}$ and $d_{\rm IU}\!=\!\sqrt{x_{\rm IU}^2+H^2}$, respectively. Under the LoS channel model, the channel from the AP to IRS,  denoted by $\boldsymbol{h}_{\rm AI}\in\mathbb{C}^{N\times 1}$, can be modeled as $\boldsymbol{h}_{\rm AI}=h_{\rm AI} \boldsymbol{a}_{\rm r}(\theta_{\rm AI}^{\rm r},\vartheta_{\rm AI}^{\rm r},N)$,  where $h_{\rm AI}\triangleq\sqrt{\beta/d_{\rm AI}^2} e^{-\jmath \frac{2\pi}{\lambda}d_{\rm AI}}$ denotes the complex channel gain with $\lambda$ and $\beta$ respectively denoting the carrier  wavelength and  the reference channel power gain at a distance of $1$ meter (m); and $\theta_{\rm AI}^{\rm r} (\vartheta_{\rm AI}^{\rm r}) \in[0,\pi]$ represents the azimuth (elevation) angle-of-arrival (AoA) at the IRS. In addition,  $\boldsymbol{a}_{\rm r}(\theta_{\rm AI}^{\rm r},\vartheta_{\rm AI}^{\rm r},N)$  denotes the receive response vector, which is given by $\boldsymbol{a}_{\rm r}(\theta_{\rm AI}^{\rm r},\vartheta_{\rm AI}^{\rm r},N)=\boldsymbol{u}(\frac{2d_{\rm I}}{\lambda}\cos(\theta_{\rm AI}^{\rm r})\sin(\vartheta_{\rm AI}^{\rm r}), N_{\rm x})\otimes \boldsymbol{u}(\frac{2d_{\rm I}}{\lambda}\sin(\theta_{\rm AI}^{\rm r})\sin(\vartheta_{\rm AI}^{\rm r}), N_{\rm y})$. Therein,  $\boldsymbol{u}$ represents the steering vector function, which is defined as $\boldsymbol{u}(\varsigma, M)\!=\![1, \!e^{-\jmath\pi \varsigma}, \dots, \!e^{-(M-1)\jmath\pi \varsigma} ]^T$, $d_{\rm I}$ denotes the distance between adjacent reflecting elements, and $N_{\rm x}$ and $N_{\rm y}$ respectively denote the number of reflecting elements along the $x$- and $y$-axis. Similarly, the LoS channel from the IRS to user, denoted by $\boldsymbol{h}_{\rm IU}^H\in\mathbb{C}^{1\times N}$, can be modeled. 

Let $\boldsymbol{\Psi}=\diag(\eta_1 e^{j\phi_1},\cdots, \eta_N  e^{j\phi_N})$ denote the reflection matrix of the active/passive IRS, where $\eta_n$ and $\phi_n$ denote respectively the reflection amplitude and phase-shift at element $n\in\{1, \cdots, N\}\triangleq\mathcal{N}$. In particular, for passive IRS, we assume the same (maximum) reflection amplitude for all reflecting elements, i.e.,  $\eta_n=1, \forall n\in\mathcal{N}$; while for active IRS, the reflection amplitude (also called the \emph{amplification factor}) of each reflecting element, $\eta_n$, is \emph{no smaller than one} due to the active reflection-type amplifier integrated. For the ease of analysis, we assume that all reflecting elements at the active IRS employ the  common amplification factor, i.e., $\eta_n=\eta, \forall n\in\mathcal{N}$.{\footnote{{\color{black}It can be easily shown that under the considered LoS channel model, all reflecting elements of the active IRS should adopt a common amplification factor due to the same channel power gain of the element-wise channels. Nevertheless, when considering the general channel model (e.g., Rician fading channel), the active IRS in general should adopt different amplification factors at different reflecting elements to achieve optimal rate performance, but at a higher implementation cost.} }} As such, $\boldsymbol{\Psi}$ can be equivalently represented as $\boldsymbol{\Psi}=\eta \boldsymbol{\Phi}$, where $\boldsymbol{\Phi}\triangleq\diag( e^{j\phi_1},\cdots,  e^{j\phi_N})$. Moreover, different from the passive IRS that reflects signals without incurring amplification noise, the active IRS generates non-negligible amplification noise at all reflecting elements, which is denoted by $\boldsymbol{n}_{\rm F}\in\mathbb{C}^{N\times 1}$ and assumed following the independent circularly symmetric complex Gaussian distribution, i.e., $\boldsymbol{n}_{\rm F}\sim \mathcal{CN}(\boldsymbol{0}_{N}, \sigma^2_{\rm F}{\bf I}_{N})$ with $\sigma^2_{\rm F}$ denoting the amplification noise power.

\section{Downlink Communication}
In this section, we consider the downlink communication and optimize the IRS placement for both the active- and passive-IRS aided wireless systems for rate maximization.\footnote{The obtained results for downlink communication can be straightforwardly extended to the case of uplink communication, which is thus omitted. }  

\vspace{-5pt}
\subsection{Active IRS}\label{Sec:DLactIRS}

Let $P_{\rm A}$ denote the AP's transmit power. For the active-IRS aided system, the received signal at the user is given by 
\begin{equation}
y_{\rm act} = \boldsymbol{h}_{\rm IU}^H \eta \boldsymbol{\Theta}( \boldsymbol{h}_{\rm AI} s +  \boldsymbol{n}_{\rm F}) + n,
\end{equation}
where $s$ denotes the  transmitted signal with power $P_{\rm A}$,  and $n$ denotes the received noise at the user with power $\sigma^2$. As such, the downlink achievable rate of the active-IRS aided system in bits per second per Hertz (bps/Hz) is given by 
\begin{equation}
R^{(\rm DL)}_{\rm act}=\log_2\l(1+\frac{ P_{\rm A}| \boldsymbol{h}_{\rm IU}^H \eta \boldsymbol{\Theta} \boldsymbol{h}_{\rm AI} |^2 }{\Vert \boldsymbol{h}_{\rm IU}^H \eta \boldsymbol{\Theta} \Vert^2 \sigma_{\rm F} ^2 + \sigma^2}\r).
\end{equation}
Note that the active IRS amplifies both the received signal and noise at each reflecting element. {\color{black}Let $P_{\rm F}$ denote the maximum amplification power of the active IRS, which is practically much smaller than the conventional RF amplifier due to the limited amplification power gain \cite{amato2018tunneling,lonvcar2020challenges}.}  Then we have 
\begin{equation}
\eta^2 ( P_{\rm A} \Vert  \boldsymbol{\Theta} \boldsymbol{h}_{\rm AI}\Vert^2+\sigma_{\rm F}^2 \Vert  \boldsymbol{\Theta} {\bf I}_{N}\Vert^2) \le P_{\rm F}.\label{Eq:powerCons}
\end{equation}
{\color{black}Moreover, to ensure that the active IRS operates at the signal-amplification mode, we should properly deploy the IRS	 such that it satisfies $\eta\ge 1$ \cite{amato2018tunneling,lonvcar2020challenges}, which indicates that the power of the incident signal at the active IRS cannot be too large (i.e., below the maximum amplification power of the active IRS in practice).}

Our objective is to maximize the downlink achievable rate for the active-IRS aided system by optimizing the IRS  phase-shift matrix $\boldsymbol{\Theta}$, IRS (common) amplification factor $\eta$, and the IRS placement parameterized by $\{x_{\rm AI}, x_{\rm IU}\}$, which is formulated as follows.
\begin{subequations}
\begin{align}
\!\!\!\!({\bf P1}):\!\max_{\substack{\boldsymbol{\Theta}, \eta, x_{\rm AI}, x_{\rm IU}} }  &\log_2\l(1+\frac{ P_{\rm A}| \boldsymbol{h}_{\rm IU}^H \eta \boldsymbol{\Theta} \boldsymbol{h}_{\rm AI} |^2 }{\Vert \boldsymbol{h}_{\rm IU}^H \eta \boldsymbol{\Theta} \Vert^2 \sigma_{\rm F} ^2 + \sigma^2}\r)
\nn\\
\text{s.t.}~~~~
& |[\boldsymbol{\Theta}]_n|=1,~~\forall n\in\mathcal{N}, \label{Eq:P1:1}\\
& x_{\rm AI}+ x_{\rm IU}=D,\label{Eq:P1:2}\\
& x_{\rm AI}\ge0, x_{\rm IU}\ge0,\label{Eq:P1:3}\\
&\eta\ge1,\label{Eq:P1:4}\\
&\! \eta^2 ( P_{\rm A} \Vert  \boldsymbol{\Theta} \boldsymbol{h}_{\rm AI}\Vert^2+\sigma_{\rm F}^2 \Vert  \boldsymbol{\Theta} {\bf I}_{N}\Vert^2) \le P_{\rm F},\label{Eq:P1power}
\end{align}
\end{subequations}
where $\boldsymbol{h}_{\rm AI}$ and $\boldsymbol{h}_{\rm IU}$ are determined by $\{x_{\rm AI}, x_{\rm IU}\}$.
First, it can be verified that problem (P$1$) is feasible if and only if $P_{\rm F}\ge NP_{\rm A}\beta/(D^2+H^2)+N\sigma_{\rm F}^2$. Next, it can be shown that 1) for any $\{\eta, x_{\rm AI}, x_{\rm IU}\}$, the optimal IRS phase-shift matrix $\boldsymbol{\Theta}$ should align the cascaded AP-IRS-user channel, i.e., $[{\boldsymbol{\Theta}}]_n=e^{-\jmath(\angle[\boldsymbol{h}_{\rm IU}^H]_n+\angle[\boldsymbol{h}_{\rm AI}]_n)}$ \cite{Wu2019TWC}; and 2) at the optimal solution, the power constraint in \eqref{Eq:P1power} is always active.
Based on the above, problem (P$1$) can be equivalently transformed into the following form for  the receive-SNR maximization by substituting the optimal IRS phase-shift.
\begin{subequations}
\begin{align}
({\bf P2}):~~\max_{ \eta, x_{\rm AI}, x_{\rm IU}} ~~ &\frac{\eta^2 P_{\rm A} \beta^2 N^2/(d_{\rm AI}^2 d_{\rm IU}^2) }{\eta^2 \sigma_{\rm F}^2 \beta N  /d_{\rm IU}^2  + \sigma^2}
\nn\\
\text{s.t.}~~~~~
&\eqref{Eq:P1:2}-\eqref{Eq:P1:4},\nn\\
& \eta^2 N(P_{\rm A}\beta/d_{\rm AI}^2+\sigma_{\rm F}^2)= P_{\rm F},\label{Eq:P2power}
\end{align}
\end{subequations}
where $ d_{\rm AI}^2= x_{\rm AI}^2+H^2$ and  $d_{\rm IU}^2= x_{\rm IU}^2+H^2$.
By substituting $\eta$ in \eqref{Eq:P2power} into the objective of problem (P$2$) and after some mathematical manipulations, problem (P$2$) can be equivalently transformed into
\begin{subequations}
\begin{align}
({\bf P3}):\max_{x_{\rm AI}, x_{\rm IU}}~~  & \frac{P_{\rm A} \beta^2 N}{C_1 d_{\rm AI}^2 + C_2 d_{\rm IU}^2 + C_3 d_{\rm AI}^2 d_{\rm IU}^2}
\nn\\
~~~~\text{s.t.}~~~
&\eqref{Eq:P1:2},\nn\\
& x_{\rm AI}\ge x_0, x_{\rm IU}\ge0, \label{Eq:P3:2}
\end{align}
\end{subequations}
where $C_1=\beta  \sigma_{\rm F}^2$, $C_2=P_{\rm A}\beta  \sigma^2 /P_{\rm F}$, $C_3=\sigma^2 \sigma^2_{\rm F}/{P_{\rm F}}$, and $x_0=\sqrt{\max\{0, NP_{\rm A} \beta/(P_{\rm F}-N\sigma_{\rm F}^2)-H^2\}}$.

Although the optimal solution to problem (P$3$) is difficult to be characterized in closed form, it can be numerically obtained by applying the one-dimensional search over $x_{\rm AI}\in[x_0,D]$. In the following, we first characterize 
 the effects of the amplification power and the number of IRS reflecting elements on the  optimal active-IRS placement.

\begin{lemma}\label{Lem:p3}\emph{The optimal AP-IRS horizontal distance to problem (P$3$), i.e., $x_{\rm AI}^*$,   is monotonically decreasing with  $P_{\rm F}$ and non-decreasing with $N$.
}
\end{lemma}
\begin{proof}
First, it can be obtained that  $x_0$ is non-increasing with $P_{\rm F}$. Next, let  $\bar{x}_{\rm AI}$ denote the optimal solution to the relaxed problem of (P$3$) by dropping the constraint  $x_{\rm AI}\ge x_0$. It is observed that $C_2$ and $C_3$ both monotonically decrease with $P_{\rm F}$. By contradiction, it can be shown that  $\bar{x}_{\rm AI}$ should be decreased to maximize its objective. Combining the above, it can be shown that $x^*_{\rm AI}$ monotonically decreases with $P_{\rm F}$. As for $N$, it is observed that  $\bar{x}_{\rm AI}$ is independent of $N$ and  $x_0$ is non-decreasing with $N$, thus leading to the desired result. 
\end{proof}
Lemma~\ref{Lem:p3} is intuitively expected since given a smaller amplification power, the active IRS should be deployed farther away from the transmitter such that it can provide a  higher amplification factor with $\eta>1$.

Moreover, to draw useful insights into the active-IRS placement design, we propose a suboptimal IRS placement design with a simple closed-form expression for $x_{\rm AI}$. To this end, we first present a useful lemma as follows.
\begin{lemma}\label{LemmaApproxSub}\emph{Given $H\ll D$, we have $\max\l\{C_1 d_{\rm AI}^2, C_2 d_{\rm IU}^2\r\} \gg C_3 d_{\rm AI}^2 d_{\rm IU}^2,\forall x_{\rm AI}\in[0,D]$, if 
\begin{equation}
\sqrt{P_A\beta}/\sigma_{\rm F}+\sqrt{P_F\beta}/\sigma\gg D.\label{Eq:LemmaAppro}
\end{equation}
}
\end{lemma}
\begin{proof}First, it can be shown that the constraint  $\max\!\l\{C_1 d_{\rm AI}^2, C_2 d_{\rm IU}^2\r\} \!\gg\! C_3 d_{\rm AI}^2 d_{\rm IU}^2, \forall x_{\rm AI}\in[0,D]$ is equivalent to 
\vspace{-5pt}
\begin{equation}
\!\min_{0\le x_{\rm AI}\le D}\max\{f_1(x_{\rm AI}), f_2(x_{\rm AI})\}\gg 1, \label{Eq:f1f2}
\end{equation}
where 
$f_1(x_{\rm AI})\triangleq\frac{P_{\rm F}\beta}{d_{\rm IU}^2  \sigma^2}=\frac{P_{\rm F}\beta}{ ((D-x_{\rm AI})^2+H^2)\sigma^2}$, and $f_2(x_{\rm AI})\triangleq\frac{P_{\rm A}\beta}{d_{\rm AI}^2 \sigma^2_{\rm F}}=\frac{P_{\rm A}\beta}{(x_{\rm AI}^2+H^2) \sigma^2_{\rm F}}.$
Next, it is observed that $f_1(x_{\rm AI})$  monotonically increases with $x_{\rm AI}$,   $f_2(x_{\rm AI})$ monotonically decreases with $x_{\rm AI}$, $f_1(x_{\rm AI}=0)< f_2(x_{\rm AI}=0)$, and $f_1(x_{\rm AI}=D)>f_2(x_{\rm AI}=D)$ given $D\gg H$. Thus, we have 
$\min_{0\le x_{\rm AI}\le D}\max\{f_1(x_{\rm AI}), f_2(x_{\rm AI})\}=f_1(\check{x}_{\rm AI}),$
where $\check{x}_{\rm AI}$ satisfies $f_1(\check{x}_{\rm AI})=f_2(\check{x}_{\rm AI})$. Given $H$ is sufficiently small as compared to $D$, we have  $\check{d}_{\rm AI}\approx\check{x}_{\rm AI}$ and $\check{d}_{\rm IU}\approx\check{x}_{\rm IU}$. As such, it can be shown that  $f_1(\check{x}_{\rm AI})=f_2(\check{x}_{\rm AI})\approx \frac{(\sqrt{P_A\beta}/\sigma_{\rm F}+\sqrt{P_F\beta}/\sigma)^2}{D^2}$. Combining it with 
\eqref{Eq:f1f2} leads to the desired result.
\end{proof} 

Note that the condition in \eqref{Eq:LemmaAppro} can be practically satisfied if $P_{\rm A}$ and/or $P_{\rm F}$ is sufficiently high. Next, based on Lemma~\ref{LemmaApproxSub}, problem (P$3$) can be approximated as
\vspace{-3pt}
\begin{subequations}
\begin{align}
({\bf P4}):~~\max_{x_{\rm AI}, x_{\rm IU}}~~  & \frac{P_{\rm A} \beta^2 N}{C_1 d_{\rm AI}^2 + C_2 d_{\rm IU}^2 } \nn\\
~~~\text{s.t.}~~~~
&\eqref{Eq:P1:2},\eqref{Eq:P3:2}.\nn
\end{align}
\end{subequations}
The optimal $x_{\rm AI}$ to problem (P$4$) can be easily obtained as
\begin{equation}
\tilde{x}_{\rm AI}=\max\l\{\frac{\sigma^2 P_{\rm A}}{\sigma^2 P_{\rm A}+\sigma_{\rm F}^2 P_{\rm F}}D, x_0\r\}.\label{Eq:Subopt}
\end{equation}
Based on \eqref{Eq:Subopt},
the  maximum downlink achievable rate  for the active-IRS aided system can be approximated as follow.
\begin{lemma}\label{Lem:ActRate}\emph{For active IRS, the maximum downlink achievable rate with optimized IRS placement given in \eqref{Eq:Subopt} can be approximated as ${R}_{\rm act}^{(\rm DL)*}\approx\log_2(1+ \widetilde{{\rm SNR}}_{\rm act}^{(\rm DL)})$, where
\begin{align}
\!\!\!\widetilde{{\rm SNR}}_{\rm act}^{(\rm DL)}\!\triangleq\!\begin{cases}
\dfrac{N \beta}{D^2}\l(\dfrac{P_{\rm A} }{  \sigma_{\rm F}^2}+\dfrac{P_{\rm F}}{ \sigma^2}\r), x_0\le \dfrac{\sigma^2 P_{\rm A}D}{\sigma^2 P_{\rm A}+\sigma_{\rm F}^2 P_{\rm F}},\\
\dfrac{P_{\rm A} \beta^2 N}{C_1 x_0^2 + C_2 (D-x_0)^2}, ~~~{\rm otherwise}.\!\!\!\!\!\!\!\!\!\!
\end{cases} \!\!\!\!\!\!\!\!\!\! \label{Eq:AppSNR}
\end{align}}
\end{lemma} 
\begin{proof} Based on Lemma~\ref{LemmaApproxSub}, the receive SNR is approximated as
\vspace{-8pt}
\begin{equation}
{\rm SNR}_{\rm act}^{(\rm DL)}\approx \frac{P_{\rm A}\beta^2 N}{C_1 (\tilde{x}_{\rm AI}^2+H^2) + C_2(\tilde{x}_{\rm IU}^2+H^2)}.\label{Eq:AppSN2}
\end{equation}
 Then,  we substitute \eqref{Eq:Subopt} into \eqref{Eq:AppSN2} and drop the term $(C_1+C_2) H^2$ in the denominator, leading to the desired result. 
\end{proof}

{\color{black}Lemma~\ref{Lem:ActRate} shows that for the active-IRS aided system with a small $x_0$ (i.e., $x_0\le \frac{\sigma^2 P_{\rm A}D}{\sigma^2 P_{\rm A}+\sigma_{\rm F}^2 P_{\rm F}}$), the (approximated) maximum receive SNR in the downlink, $\widetilde{{\rm SNR}}_{\rm act}^{(\rm DL)}$, linearly scales with $N$ (instead of quadratic scaling order in the passive IRS case), inversely scales with $D^2$, and increases with $P_{\rm F}$ and $P_{\rm A}$. Moreover, an interesting observation is that $\widetilde{{\rm SNR}}_{\rm act}^{(\rm DL)}$ can be regarded as the receive SNR at an $N$-antenna user, for which the AP-user link is LoS and the transmit SNR is $\frac{P_{\rm A} }{\sigma_{\rm F}^2}+\frac{P_{\rm F}}{ \sigma^2}$, which is the sum of that of the AP-IRS and IRS-user links. }

\vspace{-7pt}
\subsection{Passive IRS}\label{Sec:PassIRS}
For passive IRS, the received signal at the user is given by
\begin{equation}
y_{\rm pas} = \boldsymbol{h}_{\rm IU}^H  \boldsymbol{\Theta} \boldsymbol{h}_{\rm AI} s  + n.
\end{equation}
Then, the optimization problem for maximizing the downlink achievable rate can be formulated as
\begin{subequations}
\begin{align}
({\bf P5}):~~\max_{\boldsymbol{\Theta}, x_{\rm AI}, x_{\rm IU}} ~~ &\log_2\l(1+\frac{ P_{\rm A}| \boldsymbol{h}_{\rm IU}^H  \boldsymbol{\Theta} \boldsymbol{h}_{\rm AI} |^2 }{ \sigma^2}\r)
\nn\\
~~\text{s.t.}~~~~~
&  \eqref{Eq:P1:1}-\eqref{Eq:P1:3}.\nn
\end{align}
\end{subequations}
Similarly, it can be shown that the optimal IRS phase-shift matrix is to align the cascaded AP-IRS-user channel, leading to the IRS-location dependent SNR, given by ${\rm SNR}_{\rm pas}^{(\rm DL)}=\frac{P_{\rm A} \beta^2 N^2}{d_{\rm AI}^2 d_{\rm IU}^2 \sigma^2}$.
Although the closed-form optimal solution to problem (P$5$) is complicated, it can be shown that when $H$ is small, the near-optimal $x_{\rm AI}$ to problem (P$5$) is given by $\hat{x}_{\rm AI}=0$ or $\hat{x}_{\rm AI}=D$ \cite{wu2021intelligent}, which means that the passive IRS should be deployed above either the transmitter or the receiver. With the above IRS placement, the maximum receive SNR for the passive-IRS aided system is given by 
\vspace{-3pt}
\begin{equation}
\vspace{-4pt}
{{\rm SNR}}_{\rm pas}^{\rm (DL)*}\approx\widetilde{{\rm SNR}}_{\rm pas}^{\rm (DL)}\triangleq\frac{P_{\rm A} \beta^2 N^2}{H^2 (D^2+H^2) \sigma^2}.\label{Eq:pasSNR}
\end{equation}

\vspace{-5pt}
\subsection{Active IRS versus Passive IRS}
Comparing the receive SNRs for the active- and passive-IRS aided systems given in \eqref{Eq:AppSNR} and \eqref{Eq:pasSNR} under their respectively optimized placement, we obtain the following result.
\begin{proposition}\label{Prop:actpasComp}\emph{Assuming $x_0\le \frac{\sigma^2 P_{\rm A}D}{\sigma^2 P_{\rm A}+\sigma_{\rm F}^2 P_{\rm F}}$ and under the optimized IRS placement, $\widetilde{{\rm SNR}}_{\rm pas}^{\rm (DL)}\ge \widetilde{{\rm SNR}}_{\rm act}^{\rm (DL)}$ if
\vspace{-2pt}
\begin{equation}
\vspace{-1pt}
\frac{N D^2 \beta}{H^2 (D^2+H^2)} \ge \frac{P_{\rm F}}{P_{\rm A}}+\frac{\sigma^2}{ \sigma_{\rm F}^2}.\label{Eq:actpasComp}
\end{equation} }
\end{proposition}

Proposition~\ref{Prop:actpasComp} shows that given a small $H$,  the passive IRS tends to outperform the active IRS in the downlink  receive SNR when the active-IRS amplification power is too small and/or the IRS is equipped with a large number of reflecting elements. This is expected since in the above cases, the amplification factor for each reflecting element reduces, which may not be able to compensate the amplification noise.

\section{Joint Uplink and Downlink Communication}

In this section, we optimize the IRS placement for both the active- and passive-IRS cases to maximize the weighted sum-rate of  uplink and downlink communications in each case.

\vspace{-5pt}
\subsection{Active IRS}
For the active-IRS aided system, by following the IRS reflection design in Section~\ref{Sec:DLactIRS} for both the uplink and downlink, the weighted sum-rate maximization problem under the LoS channel model can be formulated as
\begin{subequations}
\begin{align}
({\bf P6}):~~\max_{x_{\rm AI}, x_{\rm IU}} ~~ &w^{(\rm UL)} R^{(\rm UL)}_{\rm act} + w^{(\rm DL)} R^{(\rm DL)}_{\rm act} 
\nn\\
~~~~\text{s.t.}~~~~
& \eqref{Eq:P1:2},\nn\\
& x_{\rm AI}\ge x_0, x_{\rm IU}\ge x_1,
\end{align}
\end{subequations}
where $w^{(\rm UL)} +w^{(\rm DL)} =1$ with $w^{(\rm UL)}\ge 0$ and $w^{(\rm DL)}\ge 0$, $x_1=\sqrt{\max\{0, NP_{\rm U} \beta/(P_{\rm F}-N\sigma_{\rm F}^2)-H^2\}}$, $$\!\!R^{(\rm UL)}_{\rm act}\!=\!\log_2\l(1\!+\!\frac{P_{\rm U} \beta^2 N^2 }{\beta N \sigma_{\rm F}^2 {d}_{\rm IU}^2 \!+\! \frac{P_{\rm U}\beta N \sigma^2 }{P_{\rm F}} {d}_{\rm AI}^2 \!+\! \frac{N\sigma^2 \sigma^2_{\rm F}}{P_{\rm F}} {d}_{\rm AI}^2 {d}_{\rm IU}^2}\r),$$ $$R^{(\rm DL)}_{\rm act}\!=\! \log_2\l(1\!+\!\frac{P_{\rm A} \beta^2 N^2}{\beta N \sigma_{\rm F}^2 {d}_{\rm AI}^2 \!+\! \frac{P_{\rm A}\beta N \sigma^2 }{P_{\rm F}} {d}_{\rm IU}^2 \!+\! \frac{N\sigma^2 \sigma^2_{\rm F}}{P_{\rm F}} {d}_{\rm AI}^2 {d}_{\rm IU}^2}\r),$$  with $P_{\rm U}$ denoting the user's transmit power in the uplink.

Although the optimal solution to problem (P$6$) is difficult to be characterized in closed form, it can be efficiently obtained by the one-dimensional search over $x_{\rm AI}$.
Generally speaking, for a practically small active-IRS amplification power $P_{\rm F}$, the active IRS should be deployed closer to the AP in the uplink, but closer to the user in the downlink (as indicated by Lemma~\ref{Lem:p3}). As such, the
 optimal IRS placement for maximizing the weighted sum-rate of both the uplink and downlink needs to strike a balance between them and depends on their rate weights in general.
 
\vspace{-8pt}
\subsection{Passive IRS}
For passive IRS, by designing the IRS phase-shift to align the cascaded channel under the LoS channel model, the weighted sum-rate maximization problem is formulated as
\begin{subequations}
\begin{align}
({\bf P7}):~~\max_{x_{\rm AI}, x_{\rm IU}} ~~ &w^{(\rm UL)} R^{(\rm UL)}_{\rm pas} + w^{(\rm DL)} R^{(\rm DL)}_{\rm pas} 
\nn\\
~~~~\text{s.t.}~~~
&\eqref{Eq:P1:2}-\eqref{Eq:P1:3},\nn
\end{align}
\end{subequations}
where $\!\!R^{(\rm UL)}_{\rm pas}\!=\!\log_2\l(1\!+\!P_{\rm U} \beta^2 N^2/(d_{\rm AI}^2 d_{\rm IU}^2 \sigma^2)\r), R^{(\rm DL)}_{\rm pas}\!=\! \log_2\l(1\!+P_{\rm A} \beta^2 N^2/(d_{\rm AI}^2 d_{\rm IU}^2 \sigma^2)\r).$

Similarly, the optimal solution to problem (P$7$) can be obtained by an exhaustive search. To obtain useful insights, we consider the case where $H$ is small. For this case, since placing the IRS above either the AP or the user is near-optimal for both the uplink and downlink communications, it can be easily shown that the near-optimal AP-IRS horizontal distance for the weighted sum-rate is $x_{\rm AI}^*=0$ or $x_{\rm AI}^*=D$.  Thus, the corresponding maximum weighted sum-rate is given by
\vspace{-2pt}
\begin{align}
&w^{(\rm UL)} R^{(\rm UL)*}_{\rm pas} + w^{(\rm DL)} R^{(\rm DL)*}_{\rm pas}\\
\approx~&w^{(\rm UL)}\!\log_2\l(1\!+\!\frac{P_{\rm U} \beta^2 N^2}{H^2 (D^2+H^2) \sigma^2}\r)\nn\\
&\quad+w^{(\rm DL)} \log_2\l(1\!+\frac{P_{\rm A} \beta^2 N^2}{H^2 (D^2+H^2) \sigma^2}\r).
\label{Eq:pasSNR2}
\vspace{-5pt}
\end{align}

\begin{remark}[Active IRS versus Passive IRS]
\emph{Compared to the downlink (or uplink) communication only, the passive IRS is more likely to yield superior rate performance to the active IRS in the joint uplink and downlink  communication. This is because the active-IRS placement needs to balance the achievable rates in the uplink and downlink, thus suffering some rate loss in each individual link; while deploying the passive IRS above either the AP or the user is  near-optimal for both the uplink and downlink.}
\end{remark}

\vspace{-10pt}
\section{Numerical Results}

In this section, we present numerical results to compare the IRS placement and rate performance for the active- and passive-IRS aided communication systems. The simulation setup is as follows, if not specified otherwise. The IRS is equipped with $N=400$ reflecting elements and deployed at an altitude of $H=1.5$ m. {\color{black}The carrier frequency is $0.75$ GHz and thus the reference path-loss is $\beta=({\lambda}/{4\pi})^2=-30$ dB, with the carrier wavelength $\lambda=0.4$ m.} Other parameters are set as {\color{black}$D=50$ m}, $P_{\rm A}=20 $ dBm, $P_{\rm U}=15$ dBm, $\sigma^2= -80$ dBm, and $\sigma^2_{\rm F}=-70$ dBm.

\vspace{-5pt}
\subsection{Downlink Communication}

\begin{figure}[t]
\centering
\subfigure[{\color{black}The accuracy of the suboptimal  AP-IRS horizontal distance in the active-IRS case.}]{\label{FigDisAppr}
\includegraphics[height=3.75cm]{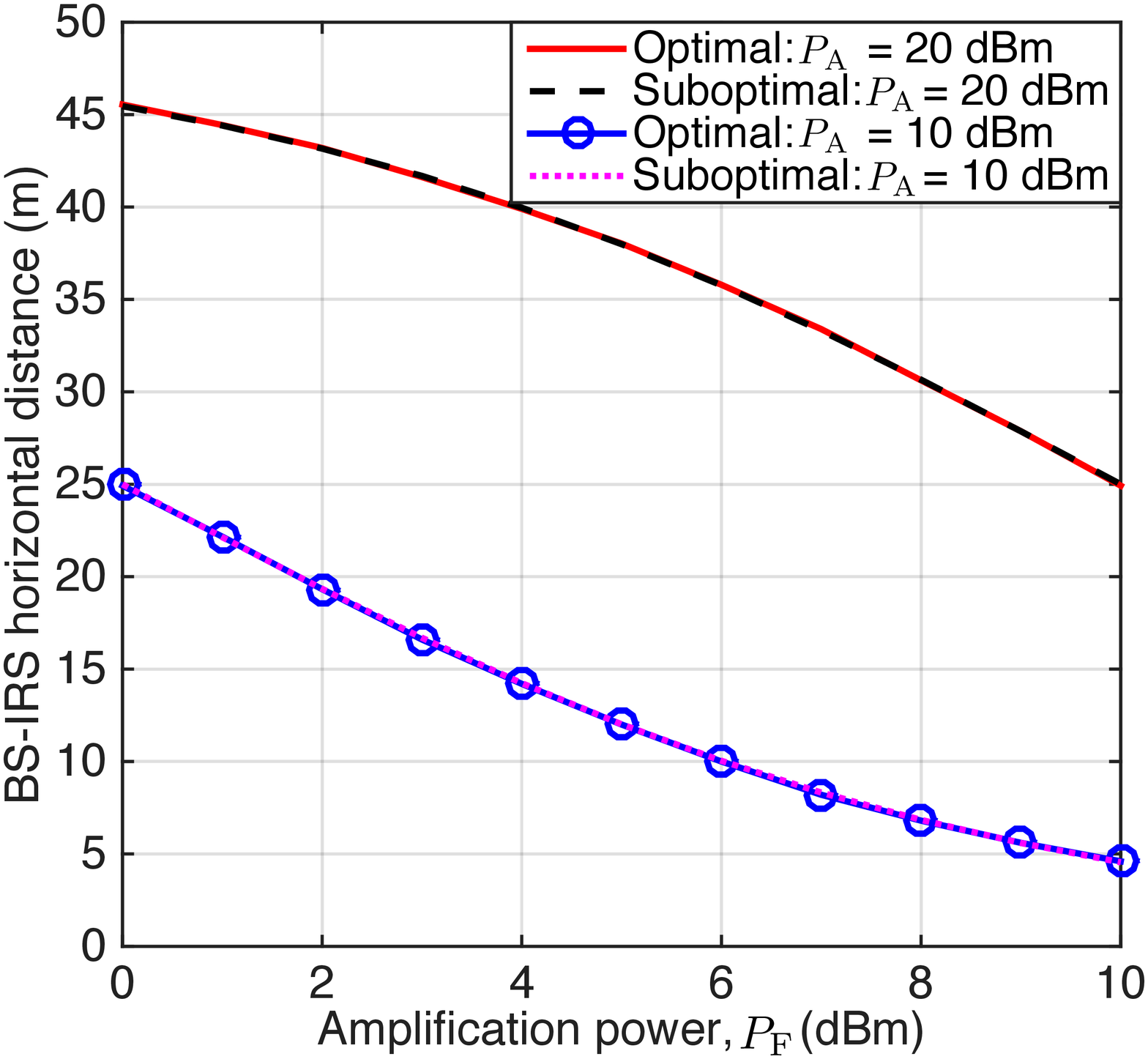}}
\hspace{1pt}
\subfigure[{\color{black}Achievable  rate versus number of IRS reflecting elements.}]{\label{FigRateNDown}
\includegraphics[height=3.75cm]{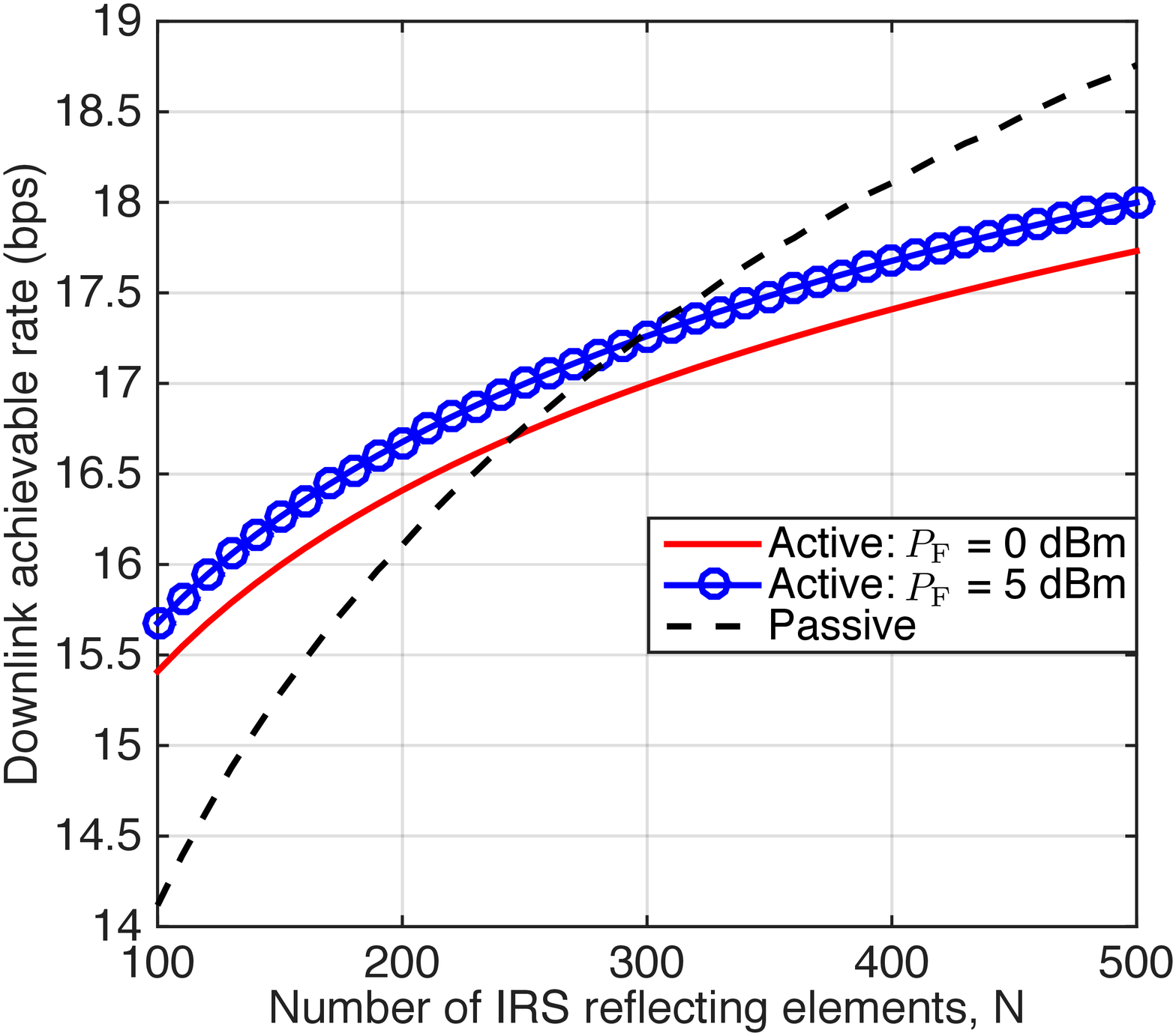}}
\label{Fig:Rate}
\caption{{\color{black}Downlink rate comparison for active and passive IRSs.}}
\end{figure}

First, we consider the downlink communication. In Fig.~\ref{FigDisAppr}, we show the optimized active-IRS placement versus the IRS amplification power $P_{\rm F}$. It is observed that for the optimal placement, the active IRS is deployed closer to the user (i.e., larger $x_{\rm AI}$) as $P_{\rm F}$ decreases. Moreover, the proposed suboptimal IRS placement is close to the optimal one, especially when $P_{\rm A}$ is large. In Fig.~\ref{FigRateNDown}, we compare the downlink achievable rates of  the active- and passive-IRS aided systems versus the number of reflecting elements $N$, where the optimal (active/passive) IRS placement is obtained by the one-dimensional search. It is observed that the downlink achievable rate of the passive-IRS case grows faster than that of the active counterpart, due to its higher power scaling order (i.e., $\mathcal{O}(N^2)$ versus $\mathcal{O}(N)$ \cite{Wu2019TWC}). Moreover, the passive-IRS aided system achieves a higher downlink rate than the active-IRS case when $N$ is sufficiently large {\color{black}(e.g., $N\ge 300$ for $P_{\rm F}=5$ dBm and $N\ge 250$ for $P_{\rm F}=0$ dBm).}

\vspace{-5pt}
\subsection{Joint Uplink and Downlink Communication}
\begin{figure}[t]
\centering
\subfigure[Weighted sum-rate versus downlink weight.]{\label{FigWeightRateW}
\includegraphics[height=3.7cm]{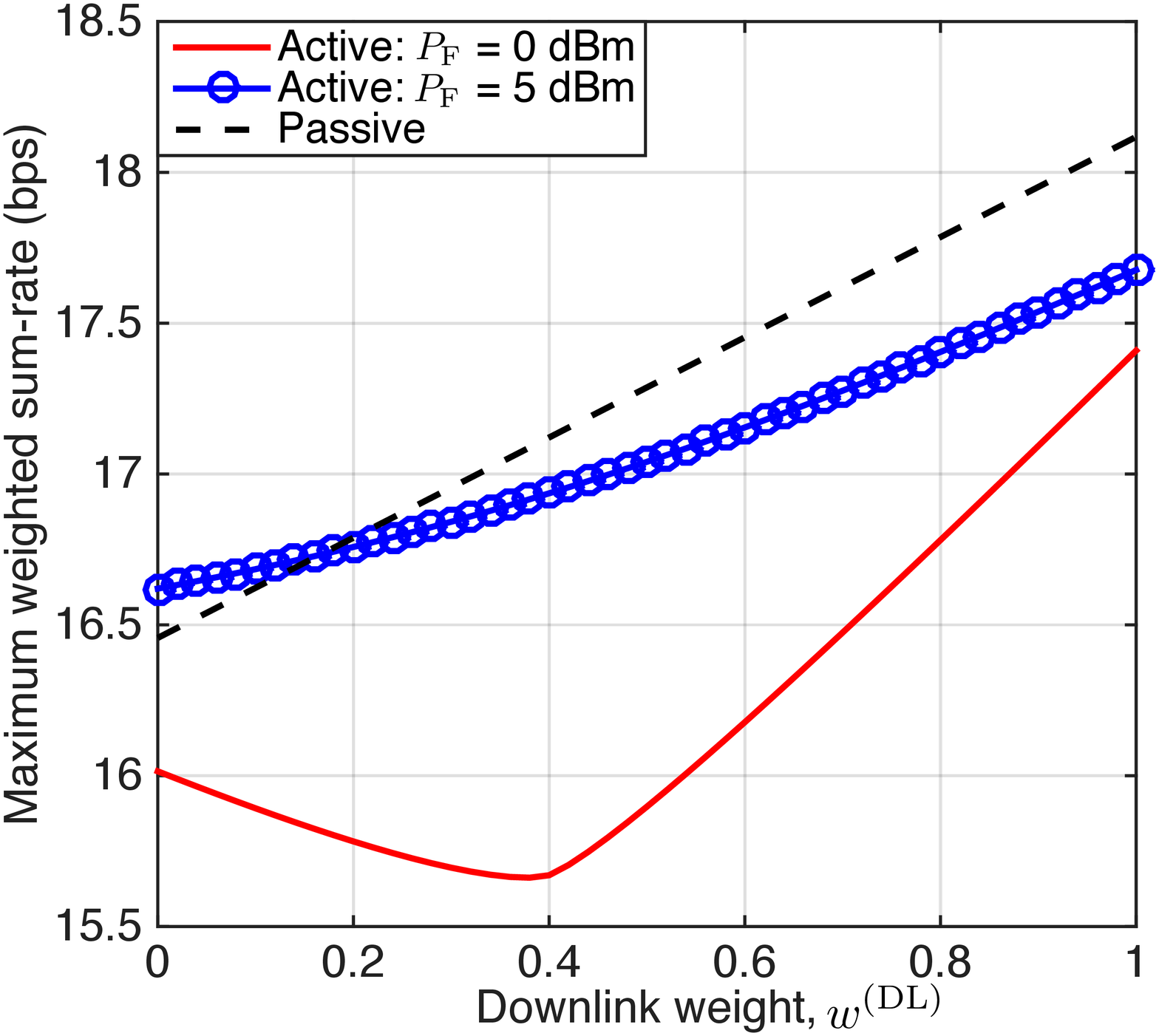}}
\hspace{2pt}
\subfigure[Weighted sum-rate versus IRS altitude.]{\label{FigRateNUpDown}
\includegraphics[height=3.7cm]{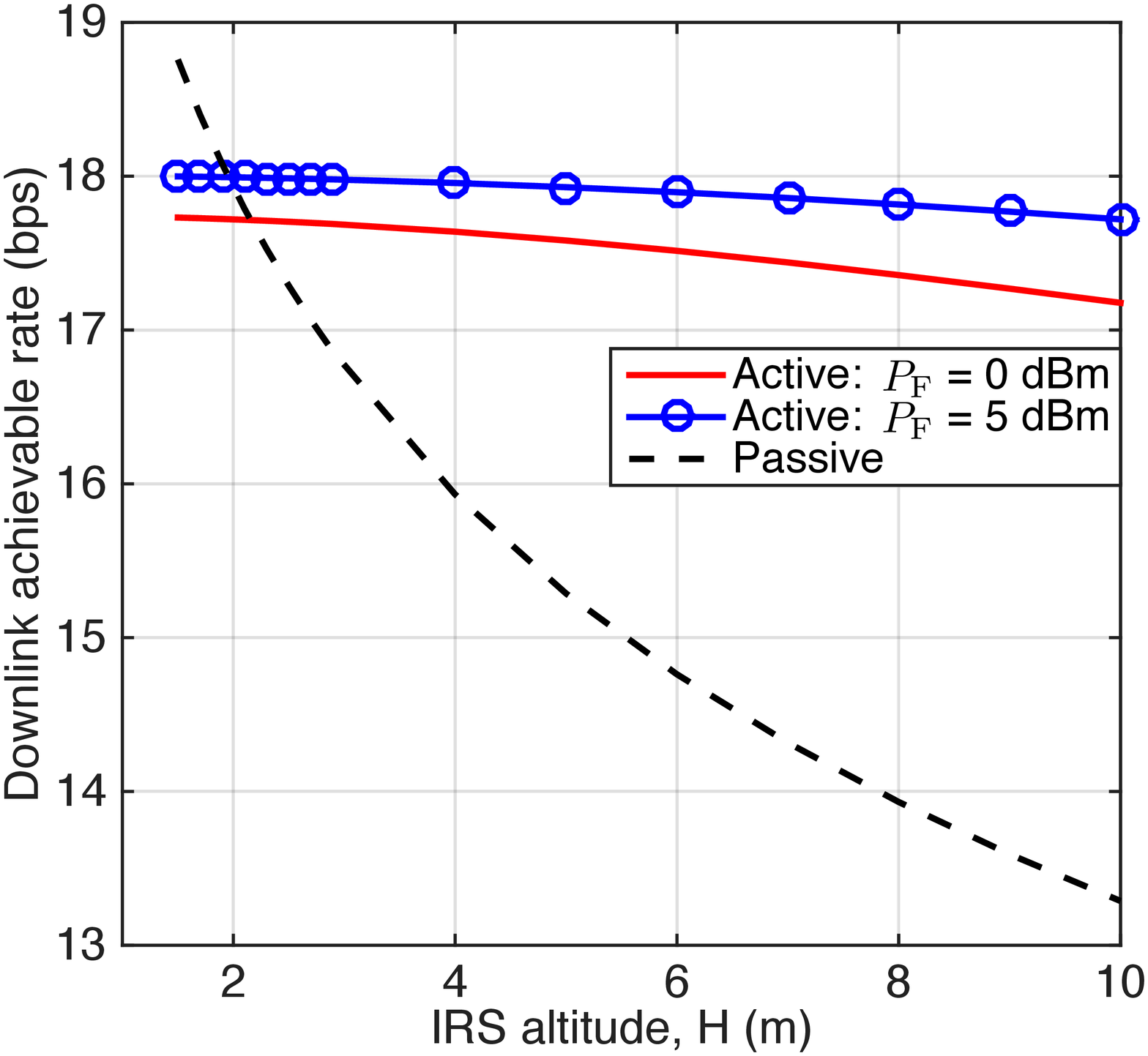}}
\label{Fig:Rate}
\caption{Weighted sum-rate comparison for active and passive IRSs.}
\end{figure}

Next, we consider the joint uplink and downlink communication. In Fig.~\ref{FigWeightRateW}, we plot the weighted sum-rate of the active- and passive-IRS aided systems under optimal IRS placement versus the downlink weight, $w^{(\rm DL)}$. First, it is observed that the weighted sum-rate of the passive-IRS case increases with $w^{(\rm DL)}$, due to  a larger transmit power at the AP than the user. While for the active-IRS case with $P_{\rm F}=0$ dBm,  its weighted sum-rate first decreases and then increases as $w^{(\rm DL)}$ increases. This is because when the uplink and downlink weights are comparable, the active-IRS placement needs to balance the rates in the uplink and downlink, thus suffering substantial rate loss in each link. Moreover, for $P_{\rm F}=5$ dBm, the rate performance gain of the passive IRS over the active counterpart  increases with $w^{(\rm DL)}$. Last, we show in Fig.~\ref{FigRateNUpDown} the weighted sum-rate of IRS aided systems versus the IRS altitude $H$, with $N=600$ and $w^{(\rm UL)}=w^{(\rm UL)}=0.5$. It is observed that increasing $H$ in general will result in substantial rate performance loss for the passive IRS, while it does not affect too much on the rate of the active IRS. This indicates that the passive and active IRSs tend to achieve superior rate performance in the regions of low and high IRS altitude, respectively. This is in accordance with the practical passive IRS deployment that should be sufficiently close to the AP.

\vspace{-5pt}
\section{Conclusions}
In this letter, we optimized the IRS placement for both the active- and passive-IRS aided wireless systems and compared their rate performance {\color{black}given the same budge on the number of IRS reflecting elements}. Specifically, for the downlink (uplink) communication, it was shown that the active IRS should be deployed closer to the receiver  with the IRS's decreasing amplification power, while the passive IRS should be deployed near  either the transmitter or receiver. With optimized IRS placement, the passive IRS was shown to achieve higher rate than the active counterpart when the number of reflecting elements is sufficiently large and/or the active-IRS amplification power is too small. Moreover, for the joint uplink and downlink communication, we showed that the passive IRS tends to achieve higher weighted sum-rate than the active IRS with respectively optimized IRS placement, because the optimal active-IRS placement needs to balance the rate performance in the uplink and downlink, while deploying the passive IRS near the transmitter or receiver is optimal for both the uplink and downlink.  {\color{black}This work can be extended in several directions for future work. For example, it is interesting to  compare the rate performance of active- and passive-IRS aided systems under other criteria (e.g, the same budge on the power consumption and/or hardware cost \cite{long2021active}) as well as under other system setups (e.g., multi-user communication).}

\vspace{-5pt}


\end{document}